\documentclass[aps,showpacs,pra,amssymb, amsmath, twocolumn]{revtex4}
\usepackage{graphicx}
\usepackage{amssymb}
\usepackage{amsmath}
\usepackage{bm}
\usepackage{xcolor} 
\begin{document}
\title{Leakage suppression by ultrafast pulse shaping}
\author{Hanlae Jo, Yunheung Song and Jaewook Ahn}
\address{Department of Physics, KAIST, Daejeon 305-701, Korea}

\begin{abstract}
We consider the leakage suppression problem of a three-level system in which the first two levels are the qubit system and the third, weakly-coupled to the second, is the leakage state. We show that phase- and amplitude-controlled two (three) pulses are sufficient for arbitrary qubit controls from the ground (an arbitrary) initial state, with leakage suppressed up to the first order of perturbation without additional pulse-area cost. A proof-of-principle experiment was performed with shaped femtosecond optical pulses and atomic rubidium showing a good agreement with the theory. \end{abstract}
\pacs{}
\keywords{quantum computation, quantum control, leakage suppression}

\maketitle

%\section{Introduction}

{\it Introduction :} Quantum two-level system stores quantum information (qubit), being the basic building block in quantum information science and engineering. However, in many physical systems, including atom and ion systems \cite{rydberg molecule, ion trap}, superconducting qubits~~\cite{DRAG_first, DRAG_EXP1}, and quantum dots~\cite{QD multi}, nature provides imperfect two-level systems, that are the two-level subspace, weakly coupled to the remaining Hilbert space of a multi-level system. Leakage transitions to unwanted energy levels are often a critical source of control errors in many physical implementations. Pulse shaping and optimal control theories have been developed to deal with the leakages and as a result to increase gate operation fidelities: there have been a plethora of model considerations for many coupled qubit systems including nuclear spins~\cite{RyanPRA2008}, superconducting qubits~\cite{SporlPRA2007,RebentrostPRA2009}, and atoms~\cite{DoriaPRL2011,TheisPRA2016}, to list a few. However, not many of them are realizable due to the current technical limit of pulse-shaping hardware. Alternatively analytic waveform designing based on physical models has been considered: one example suitable for weakly coupled nonlinear qubits is the method known as derivative removal by adiabatic gate (DRAG)~\cite{DRAG_first, DRAG_theory1, DRAG_theory2},  developed and widely adopted in superconducting qubit implementations~\cite{DRAG_EXP1, DRAG_EXP2_opt}; another is the multiple-pulse approach known as Majorana decomposition method~\cite{vitanovPRL2013, Majorana decomposition2} proposed and experimentally demonstrated recently~\cite{Majorana decomposition exp}.

Short-pulse control schemes~\cite{QD ultrafast} are often taken for quantum systems with a limited coherence time, but they could lead to rather significant leakage, especially as control bandwidth is increased comparable to intraband energies. 
In that regards, leakage suppression scheme is particularly necessary for quantum gates operating in the non-adiabatic interaction regime. For example, the time scales of ultrafast optical interactions are particularly promising for extremely fast gate operations on atomic and ion systems~\cite{ultrafast qubit op}, but both the DRAG and Majorana decomposition methods are unsuitable: the first requires too complex waveforms intractable in current technologies and the second only works for specific target states without individual coupling control.

In this paper, we consider leakage suppression by subsequent pulsed operations. We take an approach of transition pathway engineering to remove the leakage with a programmed pulse sequence. In a three-level system, it is found that an additional pulse is sufficient for leakage suppression during any qubit preparation (from the ground-state to an arbitrary state) and two pulses for any qubit rotation (from an arbitrary state to an arbitrary state). The experiments performed with cold rubidium atoms and ultrafast laser pulses show a good agreement with the prediction.

%\section{Theoretical consideration}

{\it System :} We consider a three-level system ($|0\rangle$, $|1\rangle$, and $|2\rangle$ with energies $0$, $\hbar\omega_{1}$, and $\hbar\omega_{2}$, respectively) in the ladder-type configuration, in which the first two levels are the qubit states and the third is the leakage state.
The interaction Hamiltonian $H$ (in unit of $\hbar\equiv 1$) for an electric-field $E(t)=E_0(t)\cos(\omega_L t+\phi)$ reads after the rotation wave approximation:
\begin{equation}
{H = \sum_{i = 0}^{2}{\Delta_{i}}\Pi_i+\sum_{\substack{i,j = 0 \\ i<j}}^{2}{\frac{\lambda_{ij}}{2}\left(\Omega_x~\sigma_{ij}^{(x)}-\Omega_y~\sigma_{ij}^{(y)}\right)}},
\label{eq1}
\end{equation}
where $\Pi_i = |i\rangle\langle i|$ and {$\sigma_{ij}^{(k)}$} are the identity and Pauli matrices, respectively, for $k=x,y,z$ and $i,j = 0,1,2$. $\Omega_x$ and $\Omega_y$ are the real and imaginary parts of the Rabi oscillation frequency defined by $\Omega = \mu_{01} E_0 \exp(i\phi)$ between the first and second levels, where $\mu_{01}$ is the dipole moment. Scaled dipole moments are given by $\lambda_{ij}=0$ except $\lambda_{01} = 1$ and $\lambda_{12} = \lambda$; and detunings are $\Delta_{1} = \omega_{1}-\omega_L$ and $\Delta_{2} =  \omega_{2}-2\omega_L$. 

Should the quantum information be carried  by the first two states, $|0\rangle$ and $|1\rangle$, (e.g., initial state $|\psi(t=0)\rangle = a_i|0\rangle+b_i|1\rangle$), the Hamiltonian can be divided into two parts,
the qubit system Hamiltonian $H_{q}$ and the leakage coupling Hamiltonian $H_{l}$, respectively given by 
\begin{subequations}
\begin{eqnarray}
H_{q} &=& \sum_{i = 0}^{2}{\Delta_{i}}\Pi_i+{\frac{\lambda_{01}}{2}(\Omega_x~\sigma_{01}^{(x)}-\Omega_y~\sigma_{01}^{(y)})}, \\
H_{l} &=& {\frac{\lambda_{12}}{2}(\Omega_x~\sigma_{12}^{(x)}-\Omega_y~\sigma_{12}^{(y)})}.
\end{eqnarray}
\end{subequations}
The unitary matrix for the total system dynamics can be also conveniently divided into two parts, i.e.,  $U\equiv U^{(q)}U^{(l)}$, where $U^{(q)}$ (the qubit dynamics) and $U^{(l)}$ (the leakage dynamics) are respectively governed by the following equations:
\begin{subequations}
\begin{eqnarray}
i\frac{dU^{(q)}}{dt} &=& H_{q}U^{(q)},
\label{schrodinger_boundary} \\
i\frac{dU^{(l)}}{dt} &=& U^{(q)\dagger}H_{l}U^{(q)}U^{(l)} \equiv H'_{I}U^{(l)}.
\end{eqnarray}
\end{subequations}
The resulting effective leakage Hamiltonian $H'_I$ is
\begin{eqnarray}
&H'_{I}& = \frac{\lambda}{2}(\Omega^{\ast}  U_{10}^{(q)}e^{i\Delta_2 t}\sigma_{02}^{-}+\Omega {U^{(q)}_{10}}^{\dagger} e^{-i\Delta_2 t}\sigma_{02}^{+})\nonumber\\
&+&\frac{\lambda}{2}(\Omega^{\ast} U_{11}^{(q)} e^{i\Delta_2 t}\sigma_{12}^{-}+\Omega {U^{(q)}_{11}}^{\dagger} e^{-i\Delta_2 t}\sigma_{12}^{+}).
\label{effective hamiltonian}
\end{eqnarray}
where $U_{ij}^{(q)}$ are the effective couplings defined by Eq.~\eqref{schrodinger_boundary} and $\sigma_{ij}^{\pm} = (\sigma_{ij}^{(x)} \pm i\sigma_{ij}^{(y)})/2$. So, the leakage dynamics is governed not only by the electric-field ($\lambda\Omega$) but also by the qubit dynamics ($U^{(q)}$).

When $\lambda$ is small (weakly-coupled leakage), the leakage  is obtained by eliminating the first-order perturbation terms of the Dyson series, or
\begin{equation}
c_{l} = -a_i\frac{i\lambda}{2} \int^{\infty}_{-\infty} \Omega^{\ast} U^{(q)}_{10}e^{i\Delta_2 t}dt
-b_i\frac{i\lambda}{2} \int^{\infty}_{-\infty} \Omega^{\ast} U^{(q)}_{11}e^{i\Delta_2 t}dt,
\label{first order exact}
\end{equation}
where $c_{l}$ is the leakage state coefficient after the interaction.
Then, the qubit operation fidelity~\cite{geometry ori, fidelity2, fidelity} is defined by
\begin{eqnarray}
\mathcal{F} &=& |\langle \psi(0)| U^{(q)\dagger}U^{(q)}U^{(l)}|\psi(0)\rangle|\simeq 1-\frac{1}{2}|c_{l}|^2,
%&\simeq& 1-\frac{\lambda^2}{2}|\langle\psi_i|-\frac{i}{\hbar}\int^{\infty}_{-\infty}H'_{I}dt|\psi_{\perp,i}\rangle|^2
\label{fidelity}
\end{eqnarray}
up to the first-order perturbation of the leakage.

\begin{figure*}[tbh]
\centerline{\includegraphics[width=0.80\textwidth]{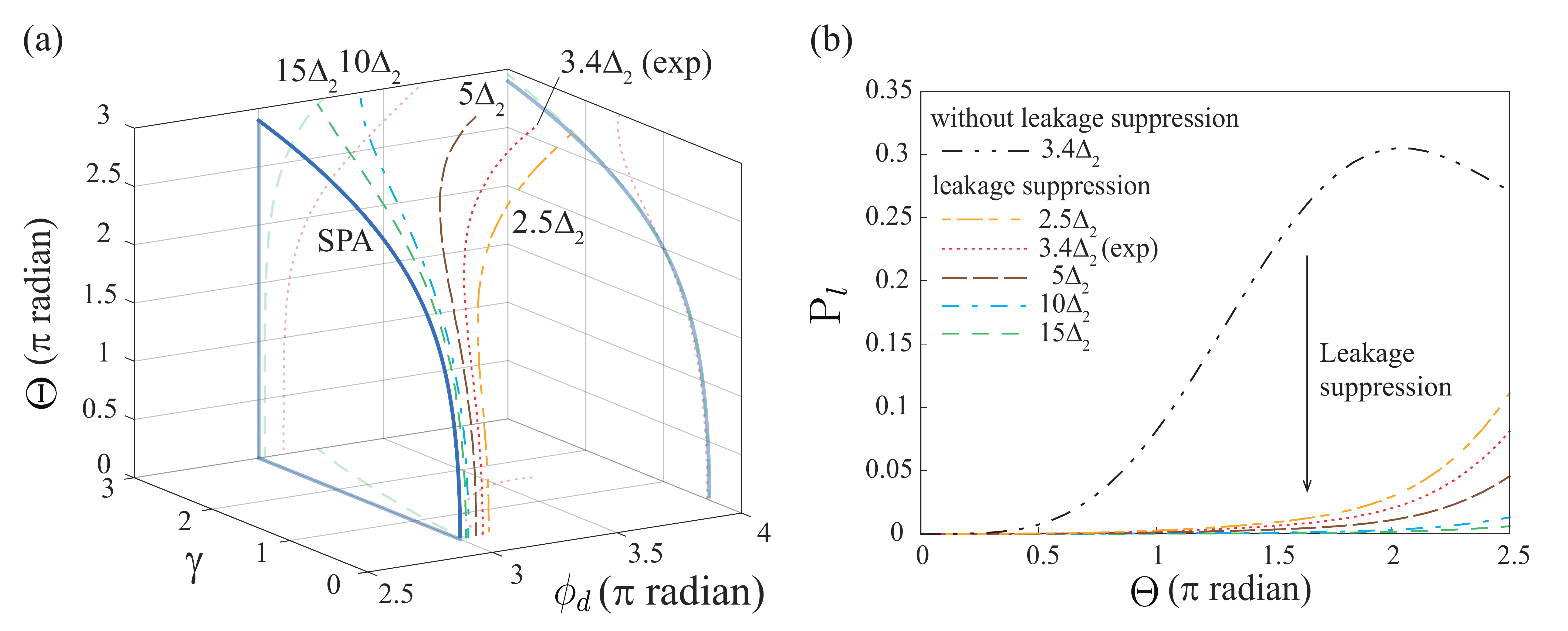}}
\caption{(Color online) {(a)} Leakage suppression condition ($c_{l}=0$) of the two-pulse scheme in Eq.~\eqref{two_pulse} for various laser bandwidths. The short-pulse approximation (SPA) solution in Eq.~\eqref{two_pulse_condition} (thick solid line) is shown in comparison with Eq.~\eqref{two_pulse} for pulses with various bandwidths ($\Delta\omega_{\rm FWHM}/\Delta_2 = 2.5, 3.4, \cdots, 15$). Our experiment corresponds to $\Delta\omega_{\rm FWHM}=3.4\Delta_2$ which is well approximated by the SPA solution up to $\Theta\sim 1.5\pi$.
(b) Leakage-state population ($P_{l}$) is compared before and after the leakage suppression, where the SPA solution in Eq.~\eqref{two_pulse_condition} is used for various laser bandwidths. The numerical calculation is performed with $\lambda = 0.34$ (atomic rubidium condition, see Experiment).}
\label{figure1}
\end{figure*}

{\it Leakage suppression with two pulses :} Now we consider  leakage suppression ($c_{l}=0$) with two pulses, which can be achieved for arbitrary final qubit-states from initial ground state (i.e., $a_i=1$). We define the two pulses by $E_1=E_0 \exp(-t^2/\tau^2)$ and $E_2=\gamma E_0 \exp(-(t-t_d)^2/\tau^2)$, with corresponding Rabi frequencies $\Omega_1$ and $\Omega_2$, respectively. When the time-dependent pulse-area is defined by $A(t) = \int^{t}_{-\infty}\mu_{12}{(1+\gamma)E_0}\exp(-t'^2/\tau^2)dt'$, 
%$A(t) = {\hbar}^{-1}\int^{t}_{-\infty}\mu_{12}{(1+\gamma)E_0}\exp(-t'^2/\tau^2)dt'$,
the corresponding time-dependent pulse-areas are given by $A(t)/(1+\gamma)$ and $\gamma A(t-t_d)/(1+\gamma)$. For the total pulse-area given by $\Theta = A(\infty)$, the target state becomes $|\psi_f\rangle = \cos {\Theta}/{2} |0\rangle-i\sin {\Theta}/{2} |1\rangle$. The carrier envelope phase (CEP) is assumed to be zero (i.e., $\phi=0$), without loss of generality. Then, Eq.~\eqref{first order exact} becomes
\begin{widetext}
\begin{eqnarray}
c_{l} = &-&\int^{\infty}_{-\infty}\frac{\lambda\Omega_1 (t)}{2} \sin{\frac{A(t)}{2(1+\gamma)}} e^{i\Delta_2 t}dt 
- \int^{\infty}_{-\infty}\frac{\lambda\Omega_2 (t_1)}{2} \cos{\frac{A(\infty)}{2(1+\gamma)}}\sin{\frac{\gamma A(t_1)}{2(1+\gamma)}}e^{i\Delta_2 (t_1+t_{d})}dt_1\nonumber\\
&-&\int^{\infty}_{-\infty}\frac{\lambda\Omega_2 (t_1)}{2} \sin{\frac{A(\infty)}{2(1+\gamma)}}\cos{\frac{\gamma A(t_1)}{2(1+\gamma)}}e^{i\Delta_2 (t_1+t_{d})}dt_1,
\label{two_pulse}
\end{eqnarray}
\end{widetext}
where $t_1 = t-t_d$. An algebraic solution for this leakage suppression ($c_{l}=0$) is then obtained as
\begin{equation}
\gamma = \left(1-\frac{2}{\Theta}\cos^{-1}{\frac{1+\cos{\frac{\Theta}{2}}}{2}}\right)\left(\frac{2}{\Theta}\cos^{-1}{\frac{1+\cos{\frac{\Theta}{2}}}{2}}\right)^{-1}
\label{two_pulse_condition}
\end{equation}
with time-delay phase {$\phi_d\equiv {\Delta_2} t_d=\pi$}, that sets the three terms in Eq.~\eqref{two_pulse} all in phase,
for short laser pulses with laser bandwidth considerably bigger than the two-photon detuning ${\Delta_2}$ under which the interaction phase ${\Delta_2} t$ is maintained constant during the interaction (short-pulse approximation). In Fig.~\ref{figure1}, the solution in Eq.~\eqref{two_pulse_condition} is compared with 
 the time-dependent Schr\"{o}dinger equation calculation  for various laser bandwidths ($\Delta\omega_{\rm FWHM}$, electric-field bandwidth), which shows that the full coverage of a Rabi half-cycle, $\Theta\in [0,\pi]$, is experimentally achievable with this leakage suppression scheme. 
   
Leakage suppression in other systems such as V-type and $\Lambda$-type three-level systems can be also considered similarly. It is found that the $\Lambda$-type systems provide the same leakage suppression condition as Eq.~\eqref{two_pulse_condition}. In the V-type systems, for example with $\lambda_{12} = 0$ but $\lambda_{02} = \lambda$ in Eq.~\eqref{eq1}, the condition is obtained in a slightly different form as 
\begin{equation}
\gamma({\rm V}) = \left(1-\frac{2}{\Theta}\sin^{-1}{\frac{\sin{\frac{\Theta}{2}}}{2}}\right)\left(\frac{2}{\Theta}\sin^{-1}{\frac{\sin{\frac{\Theta}{2}}}{2}}\right)^{-1}.
\label{two_pulse_condition_V}
\end{equation}

\begin{figure}[tbh]
\centerline{\includegraphics[width=0.5\textwidth]{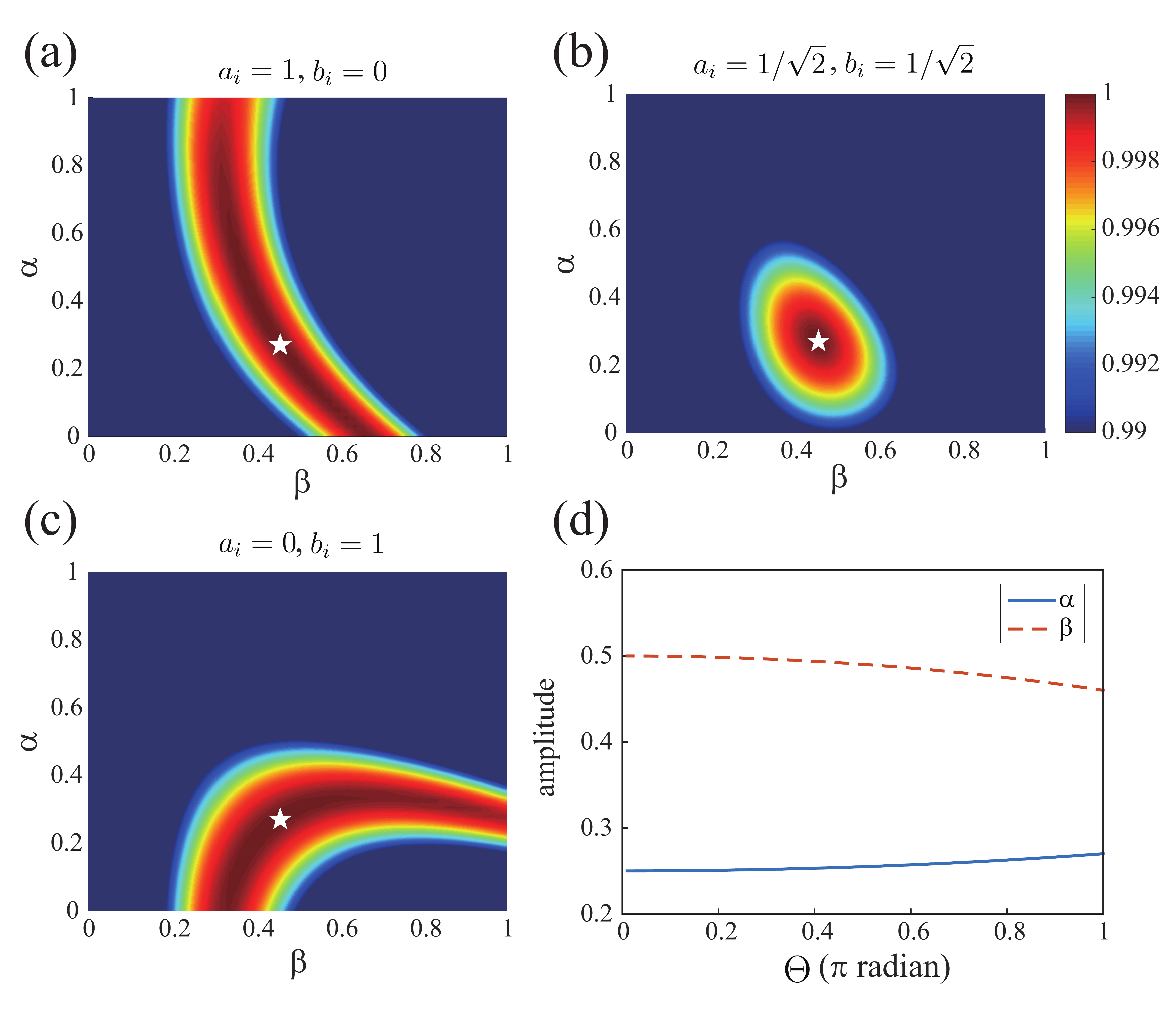}}
\caption{(Color online) (a-c) Fidelity map $\mathcal{F}(\alpha, \beta)$ for $X(\pi)$-rotation initiated from three characteristic states: (a) the ground state ($a_i = 1$, $b_i = 0$), (b) the superposition state ($a_i = 1/\sqrt{2}$, $b_i = 1/\sqrt{2}$), and (c) the excited state ($a_i = 0$, $b_i = 1$). The high fidelity regions calculated by TDSE are plotted, where white stars indicate the Eq.~\eqref{special case} solution point ($\alpha = 0.27$, $\beta = 0.46$, common in all three figures) for $X(\pi)$-rotation.
(d) The solutions of Eq.~\eqref{special case} for $(\alpha,\beta$). The TDSE calculations in (a) are performed with $\lambda = 0.34$ (atomic rubidium) and $\omega_{\rm FWHM} = 15\Delta_2$ (near the SPA condition).}
\label{figure2}
\end{figure}

{\it Leakage suppression with three pulses :} In general, qubit
gate operations require initial state independent operations. In this case, both $a_i$ and $b_i$ in Eq.~\eqref{first order exact} are nonzero, so the leakage suppression needs two additional pulses, or total three pulses. 
When the three pulses are defined by $E_1=\alpha E_0 \exp(-t^2/\tau^2)$, $E_2=\beta E_0 \exp(-(t-t_{d1})^2/\tau^2)$, and $E_3=(1-\alpha-\beta) E_0 \exp(-(t-t_{d2})^2/\tau^2)$, the leakage suppression ($c_{l}=0$) under the short pulse approximation is given by two equations:  
\begin{widetext}
\begin{subequations}
\begin{eqnarray}
(1-e^{i\phi_{d1}})\cos{\frac{\alpha\Theta}{2}}+(e^{i\phi_{d1}}-e^{i\phi_{d2}})\cos{\frac{(\alpha+\beta)\Theta}{2}}
+e^{i\phi_{d2}}\cos{\frac{\Theta}{2}} &=& 1,\\
(1-e^{i\phi_{d1}}) \sin{\frac{\alpha\Theta}{2}}+(e^{i\phi_{d1}}-e^{i\phi_{d2}})\sin{\frac{(\alpha+\beta)\Theta}{2}}
+e^{i\phi_{d2}}\sin{\frac{\Theta}{2}} &=& 0,
\end{eqnarray}
\end{subequations}
\end{widetext}
where $\alpha\Theta$ and $\beta\Theta$ are the pulse-areas; $\phi_{d1,2}$ are the phase delay due to the pulse intervals. One simple example is obtained for $\phi_{d1} = \pi$ and $\phi_{d2} = 0$ as
\begin{subequations}
\begin{eqnarray}
2\cos{\frac{\alpha\Theta}{2}}-2\cos{\frac{(\alpha+\beta)\Theta}{2}}+\cos{\frac{\Theta}{2}} = 1,\\
2\sin{\frac{\alpha\Theta}{2}}-2\sin{\frac{(\alpha+\beta)\Theta}{2}}+\sin{\frac{\Theta}{2}} = 0.
\end{eqnarray}
\label{special case}
\end{subequations}
The fidelity $\mathcal{F}(\alpha,\beta)$ after $X(\pi)$ rotation is plotted in Fig.~\ref{figure2}, respectively for three distinct initial states: (a) the ground state $(a_i, b_i)=(1, 0)$, (b) a superposition state $(1/\sqrt{2}, 1/\sqrt{2})$, and (c) the excited state $(0, 1)$. The high fidelity ($\mathcal{F}>0.99$) 
pattern on the $(\alpha, \beta)$ plane depends on initial states; however, the optimal solutions (stars in the figures) from Eq.~\eqref{special case} are the same. As a function of $\Theta$, the optimal solution is plotted in Fig.~\ref{figure2}(d).

%\section{Experimental verification}

{\it Experiment :} Experimental verification was performed with atomic rubidium 5S$_{1/2}$, 5P$_{3/2}$ and 5D states ($|0\rangle$, $|1\rangle$, and $|2\rangle$, respectively). The resonance frequency between 5S and 5P states is $\lambda = 780$~nm and between 5P$_{3/2}$ and 5D states is 776~nm, where the 5D fine-structure sub-levels are treated as one through Morris-Shore transformation{~\cite{MStransform}}. 
Experiment setup is described in our previous works~\cite{LimSR2014, LeeOL2015}. In brief, $^{87}$Rb cold atoms were prepared in a magneto optical trap (MOT) and to ensure uniform laser-atom interaction the size of the atomic cloud was kept about 80\% of laser field diameter~\cite{LeeOL2015}. Short laser pulses were produced from a Ti:sapphire mode-locked laser amplifier operating at 1~kHz repetition. The sequence of two or three pulses was programmed with an acousto-optic programmable dispersive filter (AOPDF)~\cite{AOPDF}. The center frequency was $\lambda_{\rm center} = 780$~nm and the bandwidth was $\Delta w_{\rm FWHM} = 4.5\times 10^{13}$~rad/s (corresponding to $\tau = 75$~fs), where the bandwidth was limited by the one-photon transition to 5P$_{1/2}$. The time delay between pulses was about 714~fs, corresponding to 2$\Delta t_d$ = 3.1$\pi$, and the relative phase was kept less than $\pi/10$ due to experimental imperfection. The pulse-area of the control pulse was controlled with a half-wave plate placed between a pair of cross polarizers. Probing the leakage state (5D) population was performed by ionizing the 5D state atoms with a 780~nm probe pulse after the resonance frequencies of 5S-5P$_{3/2}$ and 5P$_{3/2}$-5D were eliminated using a knife edge in the pulse compressor of the laser amplifier. The ion signal was collected by a microchannel plate (MCP). Three-photon ionization signal was subtracted from data using ion signals without the probe pulse which detected three-photon ionization only. The entire experiment was repeated at a rate of 2~Hz. 

\begin{figure}[thb]
\centerline{\includegraphics[width=0.35\textwidth]{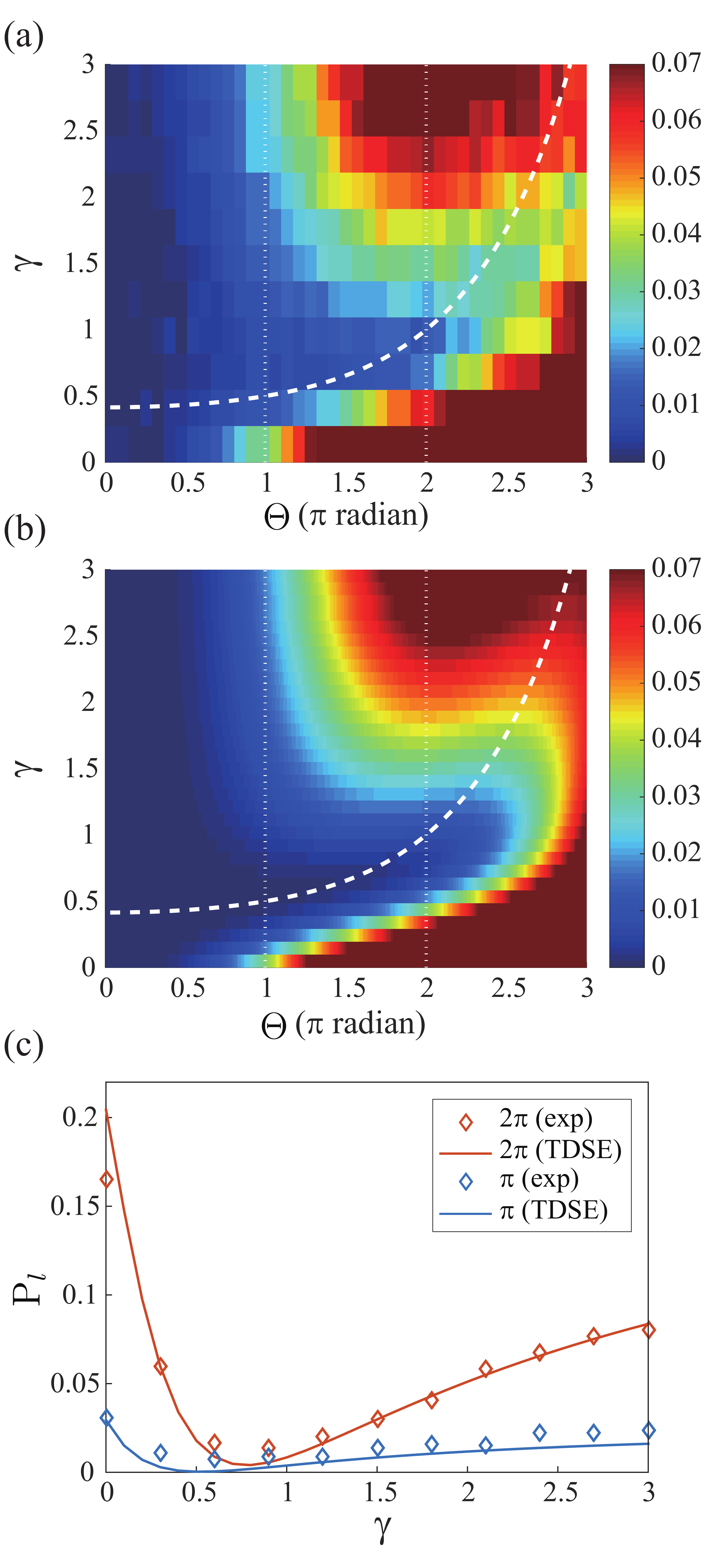}}
\caption{(Color online) (a) Experimental result, where the leakage state (5D) population is measured as a function of the total pulse area ($\Theta$) and the relative amplitude ($\gamma$). White dashed line represents the leakage suppression condition under the short-pulse approximation from Eq.~\ref{two_pulse_condition}. (b) Corresponding numerical calculation. (c) Comparison between the experiment (diamonds) and theory (solid lines) for $\Theta=\pi$ and $2\pi$, extracted along the white dotted lines in (a) and (b). 
}
\label{Experiment}
\end{figure}

The experimental result of the two-pulse leakage suppression is shown in Fig.~\ref{Experiment}(a), where the two-photon leakage to 5D state is plotted as a function of the total pulse-area ($\Theta$) and the relative amplitude ($\gamma$) between the two pulses.
In comparison, corresponding numerical TDSE calculation is plotted in Fig.~\ref{Experiment}(b). In the both figures, the dashed lines represent the leakage suppression condition in Eq.~\eqref{two_pulse_condition}, showing a qualitatively good agreement. In Fig.~\ref{Experiment}(c), the leakages ($P_l=|c_{l}|^2$) after a half ($\Theta = \pi$, blue) and full ($\Theta = 2\pi$, red) Rabi cycles are respectively plotted as a function of $\gamma$, where the data are extracted along the vertical lines in Figs.~\ref{Experiment}(a) and \ref{Experiment}(b). 

\begin{figure}[thb]
\centerline{\includegraphics[width=0.45\textwidth]{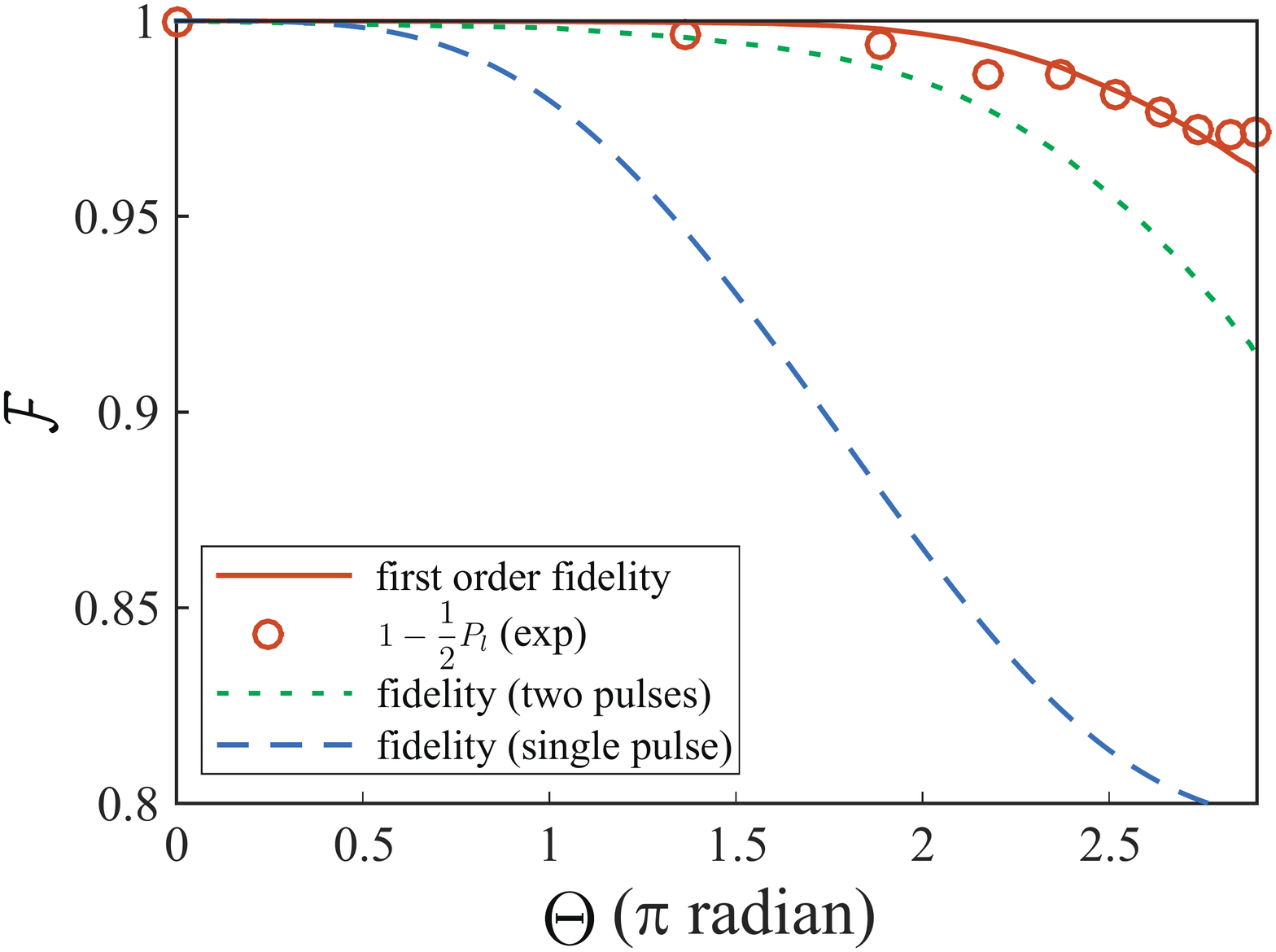}}
\caption{(Color online) Qubit-rotation fidelity $\mathcal{F}(\Theta)$. Experimental fidelity $\mathcal{F} \simeq 1-P_l/2$ in Eq.~\eqref{fidelity}, the first-order perturbation approximation, is estimated (circles) using the data from Fig.~\ref{Experiment}(a), overlapped with the corresponding theoretical estimation (solid line). In comparison, TDSE calculation for the exact fidelity
$\mathcal{F} = |\langle \psi(0)| U^{(q)\dagger}U^{(q)}U^{(l)}|\psi(0)\rangle|$ is shown for the two-pulse leakage suppression scheme (green dotted line) and for the single transform-limited pulse case (blue dashed line).}
\label{exp_fidelity}
\end{figure}

In Fig.~\ref{exp_fidelity}, the qubit rotation fidelity is estimated for our leakage suppression scheme. Using the experimentally measured leakages $P_l=|c_f|^2$ extracted along the short-pulse approximation condition (white dashed line) in Fig.~\eqref{Experiment}(a), the first-order perturbation fidelity (circles), defined in Eq.~\eqref{fidelity}, is calculated using the data extracted, and compared with the corresponding theoretical line (solid line). Direct TDSE calculations for the fidelity without the assumption of the first-order perturbation are also shown for the two-pulse scheme (green dotted line) and for the single transform-limited pulse case (blue dashed line). The estimated infidelity less than 0.025 is achieved in the $\Theta$ range from zero up to $2\pi$, using the two-pulse leakage suppression scheme.

{\it Conclusion :} We proposed a pulsed leakage suppression scheme and verified its validity using a proof-of-principle experimental demonstration. The leakage transition in a three-level atomic rubidium system was successfully removed through coherent destructive interference coherently controlled with an ultrafast pulse sequence.  The technique promises high-fidelity (exceeding $\mathcal{F} >$ 0.98) qubit rotations, requiring only phase-maintained and amplitude-controlled two pulses, for qubit rotations of a ground-state qubit, or three pulses, for general qubit rotations from an arbitrary initial state. It is expected that this coherent control scheme can be further simplified with a compressed single pulse~\cite{ZYZ rotation} and applicable not only for high fidelity quantum control but also for extremely high-speed quantum operations. 

\begin{acknowledgements}
 This research was supported by Samsung Science and Technology Foundation [SSTF-BA1301-12]. 
\end{acknowledgements}

\end{document}